\documentclass[widetext,amsmath,amssymb,twocolumn]{revtex4}
\usepackage{graphicx}
\usepackage{dcolumn}
\usepackage{bm}
\usepackage{epstopdf}
\begin{document}
\title{Robustness and Closeness Centrality for Self-Organized and Planned Cities}
\author{A Paolo Masucci}
\altaffiliation{Centre for Advanced Spatial Analysis, University College of London, 90 Tottenham Court Road, London W1N 6TR, UK }
\author{Carlos Molinero}
\altaffiliation{Centre for Advanced Spatial Analysis, University College of London, 90 Tottenham Court Road, London W1N 6TR, UK }
\date{\today}

\begin{abstract}
Street networks are important infrastructural transportation systems that cover a great part of the planet. 
It is now widely accepted that transportation properties of street networks are better understood in the interplay between the street network itself and the so called \textit{information} or \textit{dual network}, which embeds the topology of the street network's navigation system.
In this work, we present a novel robustness analysis, based on the interaction between the primal and the dual transportation layer for two large metropolises, London and Chicago, thus considering the structural differences to intentional attacks for \textit{self-organized} and planned cities. 
We elaborate the results through an accurate closeness centrality analysis in the Euclidean space and in the relationship between primal and dual space.
Interestingly enough, we find that even if the considered planar graphs display very distinct properties, the information space induce them to converge toward systems which are similar in terms of transportation properties.
\end{abstract}
\maketitle
\medskip

\section{Introduction}
Transportation systems are widely spread in nature, from biological systems, such as blood circulatory and neural systems, to ecological ones, such as ant galleries, to human mobility networks, such as train, air transport networks, etc. \cite{klein,gallos2012small,ants,cardillo2012modeling}.
Within transportation systems, street networks possibly represent the largest known infrastructural  one \cite{mikecc}.
These are growing systems whose main backbone is sometimes as old as the human civilization.
In scientific terms, street networks have been considered in graph theory and  statistical physics, where relevant statistical laws such as the Zipf's law and the Gibrat's law are still subject of a wide discussion \cite{zipf,gibrat}, while more recently, fractal theory applications  are bringing new insights into the nature of these systems \cite{mikefract,murcio2015multifractal}.

\begin{figure} [t]
\begin{center}
	\includegraphics[width=0.5\textwidth]{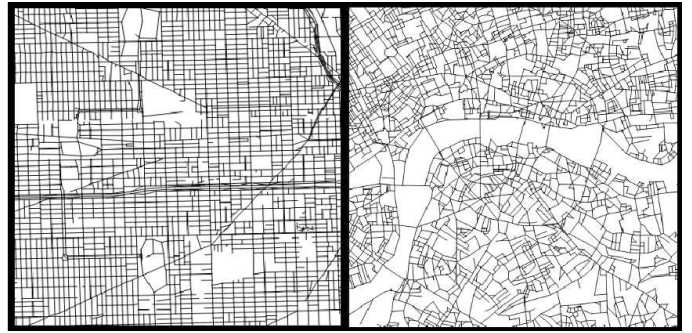}
	\caption{Examples of street network patterns extracted from the core of a planned city (Chicago in the left panel) and a self-organized city (London in the right panel).}
	\label{fig0}
\end{center} 
\end{figure}

In terms of complex systems, the street network itself, the so called \textit{primal graph}, where the nodes are the street intersections and the links are the street segments (see Fig.\ref{fig0}), does not present very interesting topological properties, with its  Poissonian-like degree distribution. 
However, with the recent  burst of activity in network theory, a quite interesting approach emerged in the field of urban studies, which is the study of street networks in its \textit{information} or \textit{dual space} \cite{rosvaldual}.
In such a representation, the vertices are collections of street segments belonging to the same transportation unit, i.e., the same road, or highway, or motorway, etc., and two vertices are linked if one ore more segments they represent intersect. 
Within this approach, it has been discovered that urban street networks display interesting properties in terms of complex systems, i.e., a fat tail connectivity distribution, which highlights hierarchical and modular behaviour, and small world properties \cite{portadual,jiangdual,pdual}. 
Such properties explain how large complex urban textures could be navigated with just a handful of information.
In this sense, it has been observed that a street network could be seen as an optimization problem tending to minimize the transport effort both in the Euclidean (also called \textit{primal space}) and the information space \cite{masuccilond}.
It is worth noting that in urban studies the dual representation is in some way similar or could be redirected in some extreme limits to \textit{space-syntax} analysis \cite{batty2004new, hillier1976space,hillier2007city,xie2007measuring}. 

\begin{figure*} [t]
\begin{center}
	\includegraphics[width=0.5\textwidth, height=0.48\textwidth]{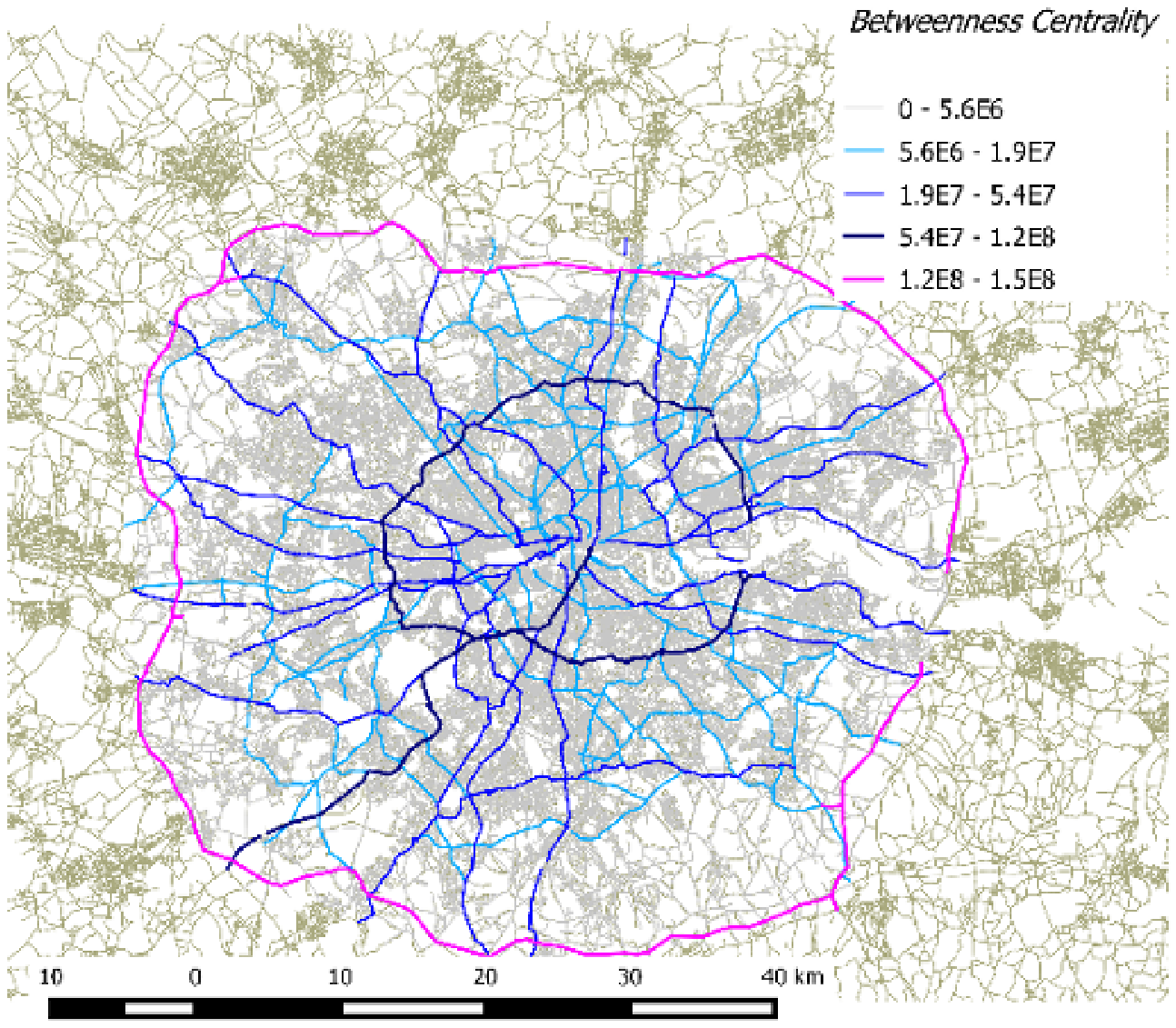}
	\includegraphics[width=0.35\textwidth]{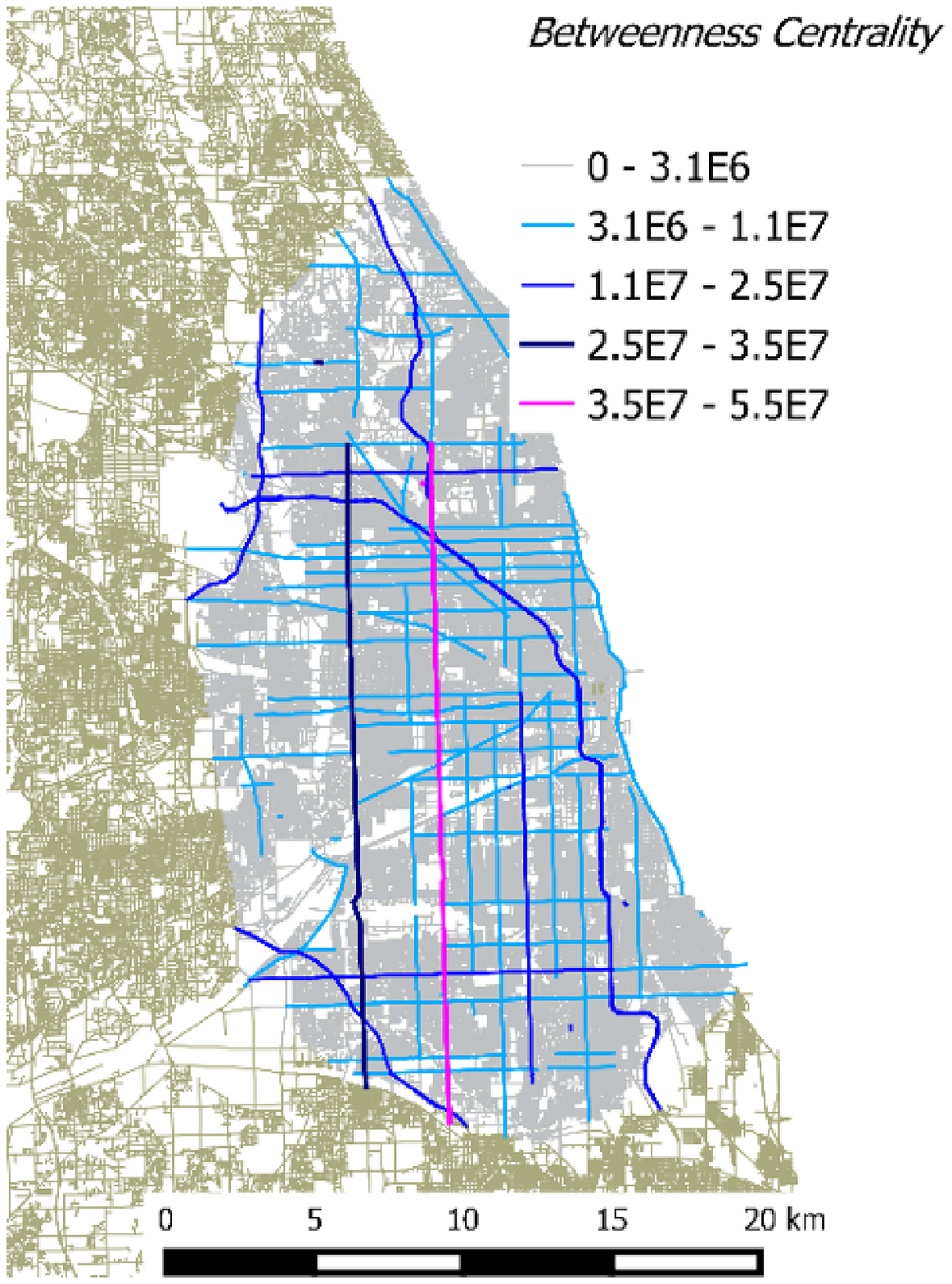}
	\caption{The dataset used in this research: in the left panel the London's street network; in the right panel the Chicago's street network. The colour map represents the betweenness centrality as calculated in the dual space and projected on the primal space.}
	\label{fig1}
\end{center} 
\end{figure*}

Street networks are not the only systems whose functions rely on double layers, where one is physical and the other one is informational. 
Between others, an interesting example is the brain cortex which shares many similarities with the street networks, as high modularity, fractal structure, etc. \cite{gallos2012small}.
Written language networks in the same way could be described by the interplay between syntactic (embedded in the text contiguity representation \cite{dorogovtsev2001language}) and  semantic layer (which relates each content word to other texts) \cite{masucci2011wikipedia}.    
In the case of street networks, we can rely on a very accurate knowledge of the primal network (even if the network extraction process is not often straight and different networks could be derived following scientist necessities or map biases),  and an always better approximation for the information space.

Only recently, it has been highlighted that in order to describe street networks, it could be important considering measures which lie in the interplay between the Euclidean and the information space \cite{viana2013simplicity, lagesse2015spatial}.
In this work, we perform an analysis of robustness for street networks   in such a spirit, considering the robustness of street networks under intentional attacks, where edges in the primal graph are chosen to be deleted, based on their betweenness centrality in their dual representation.

Using  classical techniques derived from network and information theory and introducing two novel centrality measures, we analyse the transportation properties of two large and important metropolises, London and Chicago. 
Such an exercise   helps us to shade light on some of the transportation properties of self-organised and planned cities (see fig.\ref{fig0}). 
Despite the two cities present a consistently different planar morphology, we find that the dual space is able to drive the two multi-layered systems to behave in a very similar way in terms of resilience. 

\section{Results}

\subsection{Dataset}

We consider two large metropolitan area street networks, namely London and Chicago \cite{meridian,illinois}.
This  choice is based on the idea of understanding how very different street networks morphologies could affect the transportation properties of the system.
In this sense, we choose London as the largest West-European metropolis, and as a representative of the so-called \textit{self-organized} cities, which are cities with a millenary history and whose street network grew without a single predominant urbanistic plan \cite{mikecc}.
In order to define London's boundaries, we choose to consider the street network comprised in the M25 orbital road \footnote{
Defining city boundaries is still an open debate. 
Our choice relates on the fact that we are performing an analysis about urban transportation properties. 
In this sense, choosing a natural definition for city boundaries \cite{masucci2015logistic,arcaute2015constructing,jiang2011zipf} would cut in several pieces the M25, which is an extremely important orbital road in order to navigate the city in the information and in the Euclidean space. 
}.
As  London's counter part, we choose Chicago as a representative of planned cities. 
Chicago is the third largest city in the USA in terms of population.
Chicago's street network has a relatively short history as it was incorporated as a city in 1837 and had a rapid expansion in the mid-nineteenth century, and its urban plan is mostly reticulate.
In the case of Chicago, given the lack of an entity such as an orbital road, we consider the urban area surrounding Chicago as defined by the condensation threshold methodology \cite{masucci2015logistic}.

In order to define the dual representation of these cities, we employ a hybrid hierarchical methodology.
For motorways, primal and secondary roads, we use the street name approach \cite{jiangdual}, while for minor roads we employ the hierarchical intersection continuity negotiation approach as introduced in \cite{pdual}, which is a sophistication of the intersection continuity negotiation as introduced in \cite{portadual}.

In Fig.\ref{fig1}, we show the resulting dataset. 
The color map in the figure represents the betweenness centrality of the dual representation, as projected on top of the primal representation.
We can appreciate how this measure in the dual space well reflects the main arteries of both the cities, and thus reflects the cities main navigational topology.  
   
   \begin{table*}[t]
\begin{tabular}{|c|c|c|c|c|}
\hline
 &London Primal & London Dual & Chicago Primal & Chicago Dual  \\ \hline
 N&75919 &34628  &53882  & 21244 \\ \hline
 E&107356 &	68104 & 86944 & 61010 \\ \hline
 $\langle k \rangle$&2.83 & 3.93&	3.23&	5.74  \\ \hline
 $k_{max}$&10 &286  & 9 &363  \\ \hline
Diam &209 & 14 & 289 & 19 \\ \hline 
 A[km2]&2300 &-  & 1150 &-  \\ \hline
 L[km]&15016 & - & 12294 &-  \\ \hline
 $\langle l \rangle$[m]&140 & - & 141 &-  \\ \hline
 $L_{DelTr}$[km] &40143 & - &26923  & - \\ \hline
 $L_{MST}$[km] &6479 & - &4829  & - \\ \hline
\end{tabular}
\caption{Main geometrical and topological properties of London and Chicago's street network in their primal and dual representation. The number of vertices $N$, the number of edges $E$, the average degree $\langle k \rangle$, the maximum degree $k_{max}$, the topological diameter $Diam$, the total network area $A$, the total street length $L$, the average street length  $\langle l \rangle$,  the total street length for the Delaunay triangulation $L_{DelTr}$, the total street length for the minimum spanning tree of the Delaunay triangulation $L_{MST}$.}
\label{tab1}
\end{table*}

 \subsection{Analysis}

\begin{figure} [b]
\begin{center}
	\includegraphics[width=0.5\textwidth]{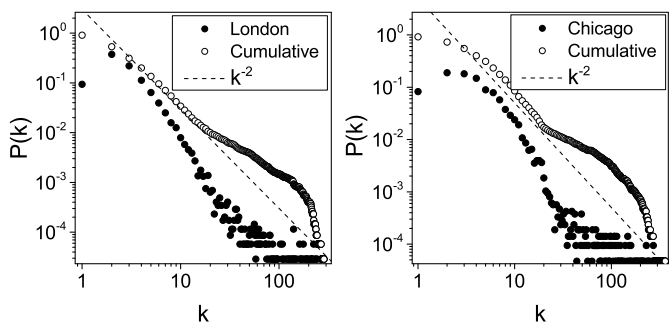}
	\caption{Degree distribution and complementary cumulative degree distribution for the dual network of London (left panel) and Chicago (right panel). The dashed line represents the curve $k^{-2}$ and it serves solely as a qualitative eye inspection tool.}
	\label{fig2}
\end{center} 
\end{figure}

In Tab.\ref{tab1}, we show some of the main geometrical and topological properties of the street networks which are considered in this research, while in Fig.\ref{fig2} we display the degree distribution together with its cumulative distribution for the networks in the information space.
As we can see from the figure, we cannot talk about scale-free distributions. 
However, it is worth to notice that we are considering just artificially delimited sections of wider street networks. 
It has already been  noticed that considering the whole networks in the information space, better results for scaling are obtained \cite{kalapala2006scale}.
For us, what it is important to notice  is that we are dealing with broad distributions characterized by a fat tail.
This implies the presence of hubs and a strong hierarchy in the dual space. 
Also, it is important to highlight the value of the diameter of these networks in the dual space  (see Tab.\ref{tab1}), whose order of magnitude is the same as the logarithm of the network sizes, thus telling us how such networks display small world properties in the information space \cite{watts1998collective}. 
These information elucidate the already known fact that the information space represents a hierarchical complex systems lying on top of the planar graph \cite{kalapala2006scale}. 

\subsubsection{Informational robustness}

Exact robustness analytical results are known for random and preferential attachment topological networks \cite{dorogovtsev2013evolution}. 
Planar graphs are always difficult to treat analytically for the planarity criteria.
Nevertheless, empirical study for street network robustness have been performed, based on the properties of the primal graph \cite{buhl2006topological}, and some analitycal results are known for a few regular lattices in percolation theory \cite{Stauffer}.
However,  in order to better understand how a street network reacts to intentional attacks, we consider the information space as its main topological characterization, analysing what we call the \textit{informational robustness}.

In order to perform the informational robustness analysis, we pick up nodes in the dual graph, with probability proportional to their betweenness centrality in that space, where the betweenness centrality for a node $k$ is defined as the number of shortest paths $n_{ij}(k)$ passing though it, i.e., $C_{BET}\equiv \sum_{i\neq j \neq k} \frac{n_{ij}(k)}{n_{ij}}$. 
Then, we consider the  road in the primal graph which corresponds to the selected node in the dual graph and we remove a randomly chosen street segment from it. 
Removing such a segment changes the topology of the dual graph, which is then updated, i.e., where the road has been broken, we assign new different ids to the two  new roads and update the whole network to this regard. 
Finally, we update the betweenness centrality for the new dual graph and start the process again.

The main idea behind the informational robustness is that important roads, where  long distance traffic in cities happens, and that give small world properties to dual urban networks, are highlighted with great accuracy by the betweenness centrality in the dual space (see Fig.\ref{fig1}).
If we want to consider intentional attacks or faults that could affect greatly urban traffic, then we choose those roads which have a large value of betweenness centrality in the dual space.   

\begin{figure} [b]
\begin{center}
	\includegraphics[width=0.48\textwidth]{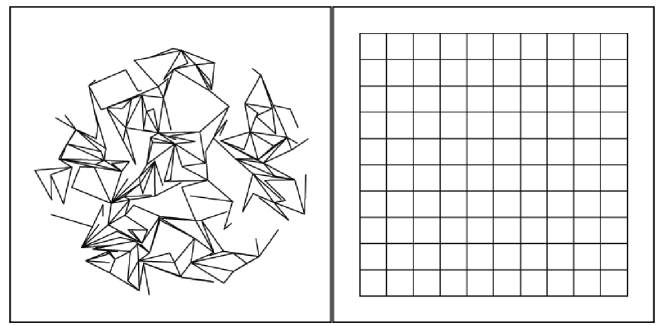}
	\caption{In the left panel an example of ERPG graph with 185 vertices and average degree $\langle k \rangle=3.24$. In the right panel an example of GRID with 121 vertices.}
	\label{fig2b}
\end{center} 
\end{figure}

To better understand the results, we introduce two null models, an Erd\"{o}s R{\'e}nyi planar graph \cite{masuccilond} (ERPG here after), and a perfect square grid (GRID hereafter) (see Fig.\ref{fig2b}, we built the networks to have 10000 vertices each).
The ERPG is built to have approximatively the same average degree of our sample cities and it is worth noticing that since it does not own a road hierarchy as the analysed street networks, the dual graph is calculated using the standard ICN method \cite{portadual}.
The GRID is a classical square lattice, whose percolation results are known analytically \cite{Stauffer}. 
Its dual representation consists of a bipartite graph, where one class of nodes are the vertical roads, and the second class of nodes are the horizontal roads \cite{masuccilond}.

Moreover, in order to frame the informational robustness analysis, we compare it to  the classical robustness analysis on the primal representation, which is done by randomly removing street segments in the planar graph. 

\begin{figure*} [t]
\begin{center}
	\includegraphics[width=0.7\textwidth]{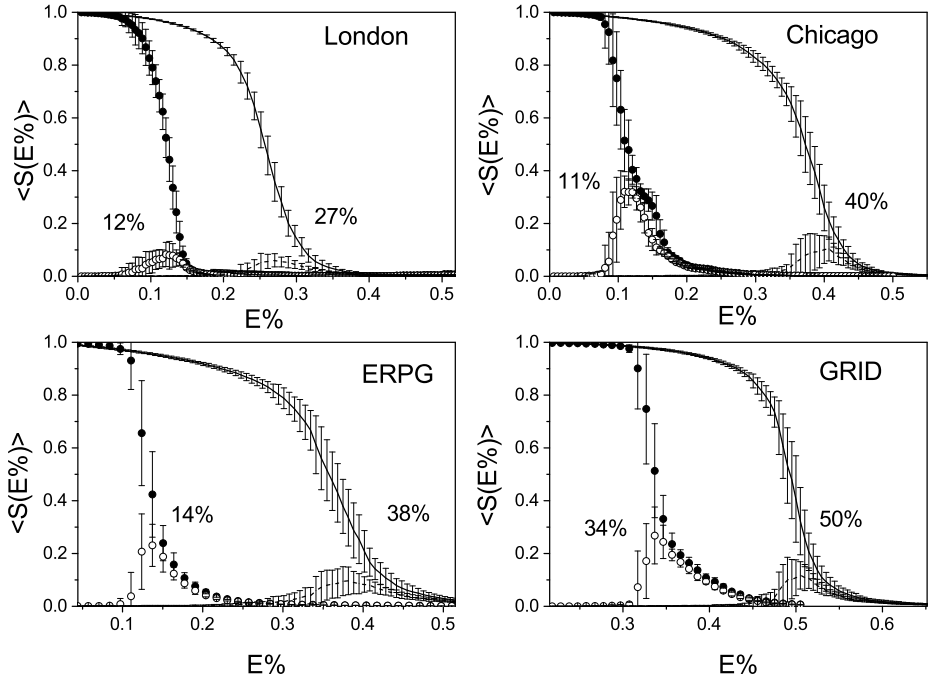}
	\caption{Robustness analysis for London (top-left panel), Chicago (top-right panel), ERPG (bottom-left panel) and GRID (bottom-right panel). Black circles: average maximum cluster size in the informational-robustness analysis; White circles: average second largest cluster size in the informational-robustness analysis; Full line: average maximum cluster size in the classical-robustness analysis; Dashed line: average second largest cluster size in the classical-robustness analysis. Error bars represent the standard deviation over 50 realizations. The numbers represent the network fracture points.}
	\label{fig3}
\end{center} 
\end{figure*}

In Fig.\ref{fig3},  we show the normalized size of the largest and second largest primal network component $\langle S(E_\%) \rangle$, averaged over 50 realizations of the process, versus the percentage of removed edges $E_\%$ in the primal network, for the informational and classical robustness analysis. 
Following classical percolation theory \cite{Stauffer}, we localize the point for the network fracture where we have a steep drop of the maximum cluster size, associated with a peak for the second largest component size.

Starting by analysing the classical robustness results in the primal space, we can see how the GRID is the most robust network between all the presented cases, with a fracture point around the $50\%$ of edge removal, as expected from percolation theory.
Such a high value for the fracture point, with respect to the other networks, depends on the fact that the GRID has a larger number of links, i.e., the average degree is 4.
After the GRID,	 we find the Chicago street network, which breaks for $E_\%\approx 40$. 
This behaves as expected, given its reticulate structure.
At the third place we find the ERPG, which breaks at $E_\%\approx 38$ and as the most fragile one, we find  London's street network which breaks at $E_\%\approx 27$.

The robustness difference  between a self-organized city as London and a planned one as Chicago in the primal space is striking, but the situation changes greatly when we calculate the informational robustness.
Still the network with the highest robustness is the GRID, which breaks after $34\%$ of edge removal. 
We can appreciate anyway how much the information space, which generates the navigation routes, could be influential in a disruption problem.
After that, we find the ERPG, which breaks at $E_\%\approx 14$.
Then we find a complete inversion of tendency,  with London at the third place, breaking at $E_\%\approx 12$, and finally Chicago, whose network breaks at $E_\%\approx 11$.

The highly modular structure of real street networks in the primal space behaves poorly in terms of robustness compared to random networks, but  we have to keep in mind  that cities are transportation systems, so that a certain degree of fragility is due to ensure best transport in the interplay between primal and dual space.
This is not surprising as the less fragile system, the GRID, is the less efficient for transportation in terms of shortest paths between nodes \cite{masuccilond}.

In fact, another way to read these results is how much the hierarchical information network, which lies upon the primal graph, is influential in the navigation of the system.
Then, we can appreciate how much urban transportation systems rely on the hierarchical information system for their functioning. 
Interestingly enough, we find that even if London and Chicago are very different in terms of primal networks, the information space let them converge toward similar navigation properties, such as a very close fracture point. 

\begin{figure} [htb]
\begin{center}
	\includegraphics[width=0.5\textwidth]{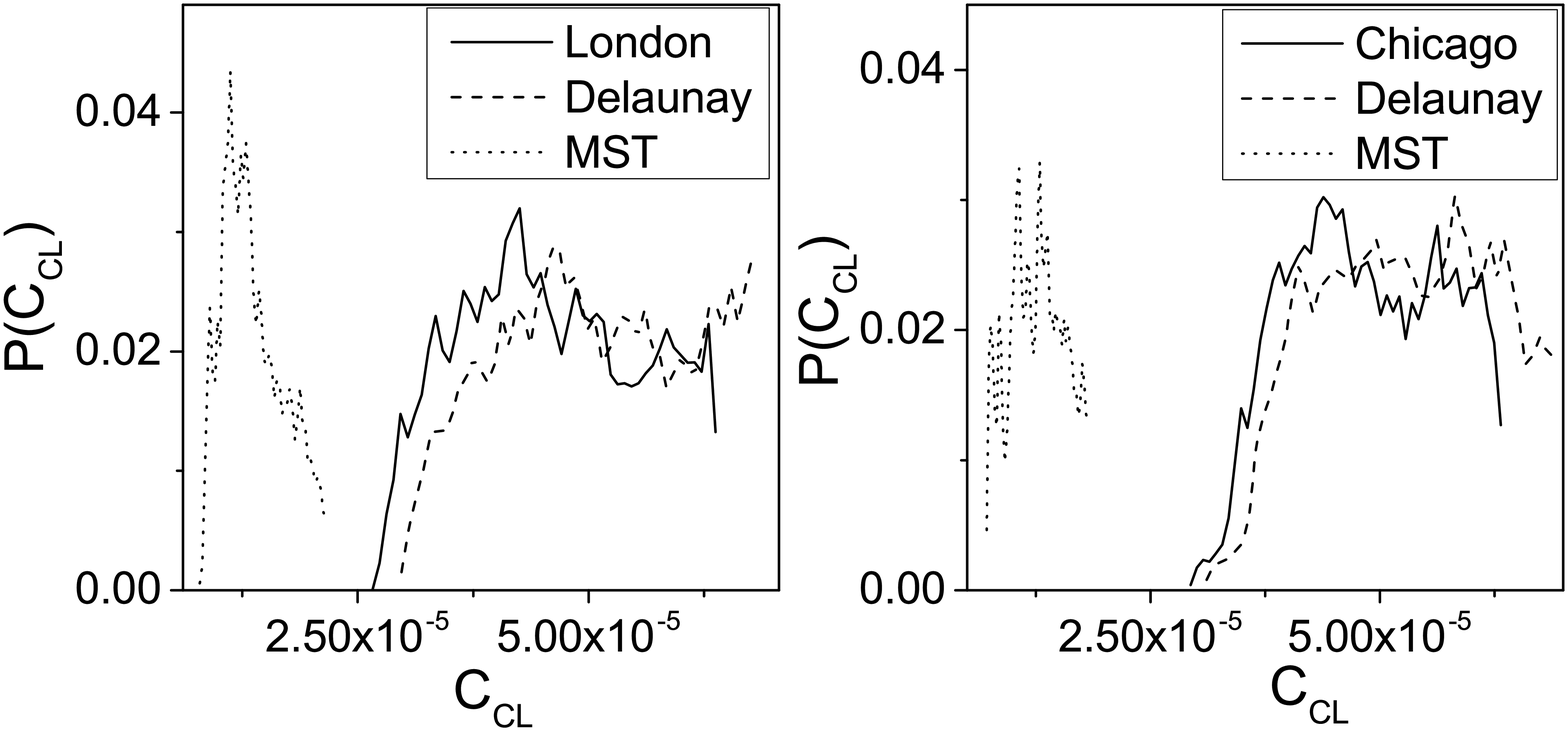}
	\includegraphics[width=0.5\textwidth]{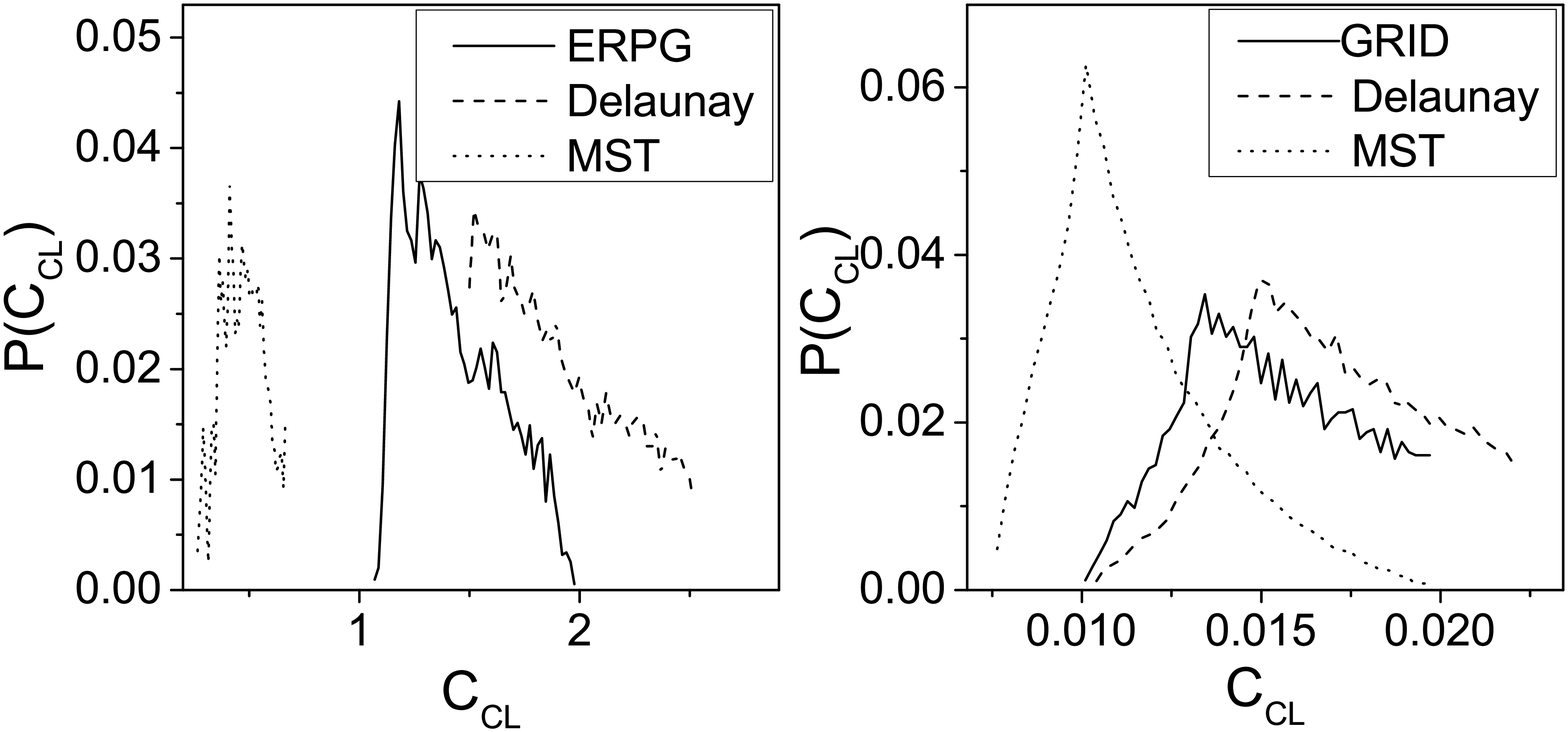}
	\caption{Closeness centrality distribution $P(C_{CL})$, calculated for the street networks and models, along with the Delaunay triangulation for their vertices, and the minimum spanning tree (MST) derived from the Delaunay triangulation: London (top-left), Chicago (top-right), ERPG (bottom-left), Grid (bottom-right).}
	\label{fig4}
\end{center} 
\end{figure}

\subsubsection{Closeness centrality}

To better understand this, we show in Fig.\ref{fig4} the closeness centrality analysis for our networks, where the closeness centrality is defined as $C_{CL}\equiv (N-1)/\sum_{i\neq j} d_{ij}$,  $d_{ij}$ is the  Euclidean distance calculated on the shortest path that connects node $i$ and node $j$, and $N$ is the number of nodes.
The closeness centrality spectrum analysis helps us to understand how the primal graph behaves locally in terms of transportation efficiency on the raw street network (in this case the betweenness analysis would give us poor results, since as we have seen the topology of a street network is mainly delivered by its information space). 

In each panel of the figure, we compare the closeness centrality distribution $P(C_{CL})$ of the selected networks, with the one derived performing a Delaunay triangulation of the network intersections, and the one derived extracting the minimum spanning tree (MST hereafter) of the Delaunay triangulation.
This comparison will allow us to speculate about the efficiency of a street network in the primal space as compared to its fragility, which we analysed before in the interplay between primal and dual space.

The Delaunay triangulation is a planar graph  connecting all the neighbouring vertices, maximising the minimum angle for all the triangles, thus creating a network where all the routes are possible between a point and another one.
Such a network represents the most efficient one in terms of travelling in the primal space, at the expenses of the total street network length (see Tab.\ref{tab1}). 
On the other hand, the MST is a network which connects all the street intersections using the minimum number, and length, of links, and thus represents the less efficient network which connects all the intersections in order to navigate the net in the primal space, but whose total street network length is minimal (see Tab.\ref{tab1}).
We use this artifice because a direct comparison between the closeness centrality distribution for the different networks is difficult to perform, due to the sensitivity of such a measure to the size of the graph.

First of all, we notice how the distributions $P(C_{CL})$ for the real systems are consistently different from the ones of the models, where after a first high peak the distribution drops steeply. 
This behaviour is understood considering that for the models the closeness centrality follows the homogeneous geometry of the network, while for the real systems the highly modular geometry of the street network ensures a more uneven distribution.

Interestingly enough, we notice how the closeness distributions for the real urban networks  are very close to  Delaunay's ones, rather than to the MST's ones. 
This is quite impressive if we look at the total length of the networks (see Tab.\ref{tab1}).
In particular, we could consider the ratio $\sigma$ between the total length of the street network and the total length of its Delaunay triangulation as an intuitive measure of the street network efficiency in the primal space.
Such a measure is not sensitive to the size of the net, as it was showed that the total street length $L$ of a urban street network is consistent with a linear function of $N$, i.e., $L(N)\propto N$ \cite{masucci2015logistic,masuccipo}.
Then we find that $\sigma_{London}\approx 0.36$, $\sigma_{Chicago}\approx 0.46$, $\sigma_{ERPG}\approx 0.65$, $\sigma_{GRID}\approx 0.59$. This means that  real street networks, at least the ones we analyse, perform in terms of efficiency in the primal space in a way that is very close to the maximum, with a length of the network that is just around  $40\%$ of their Delaunay triangulation.
In the same way, we can see also how they are optimized  with respect to the null models, whose $\sigma$ lies around the $60\%$ of the Delaunay triangulation.

\subsubsection{Information content and informational closeness centrality}

\begin{figure} [t]
\begin{center}
	\includegraphics[width=0.5\textwidth]{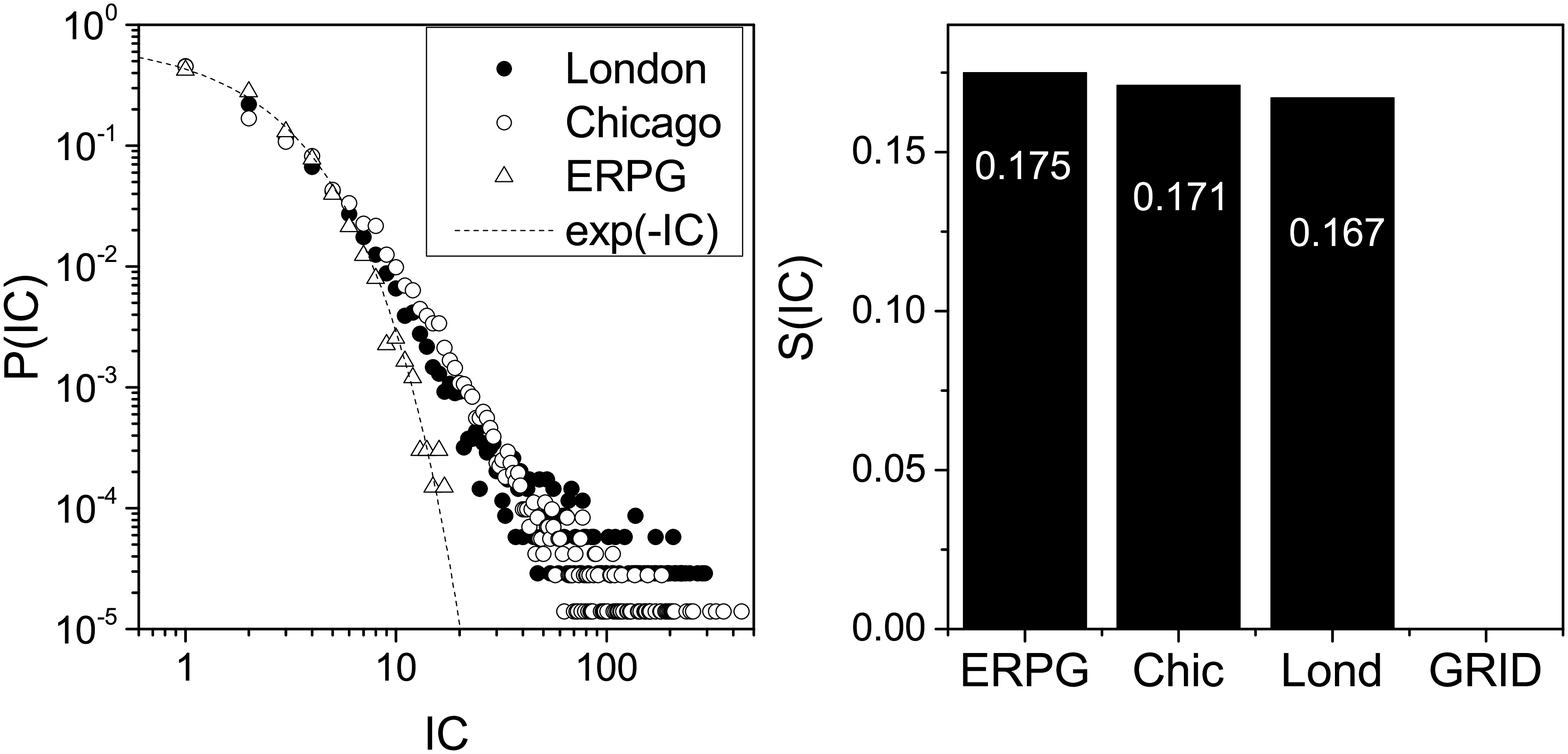}
	\includegraphics[width=0.5\textwidth]{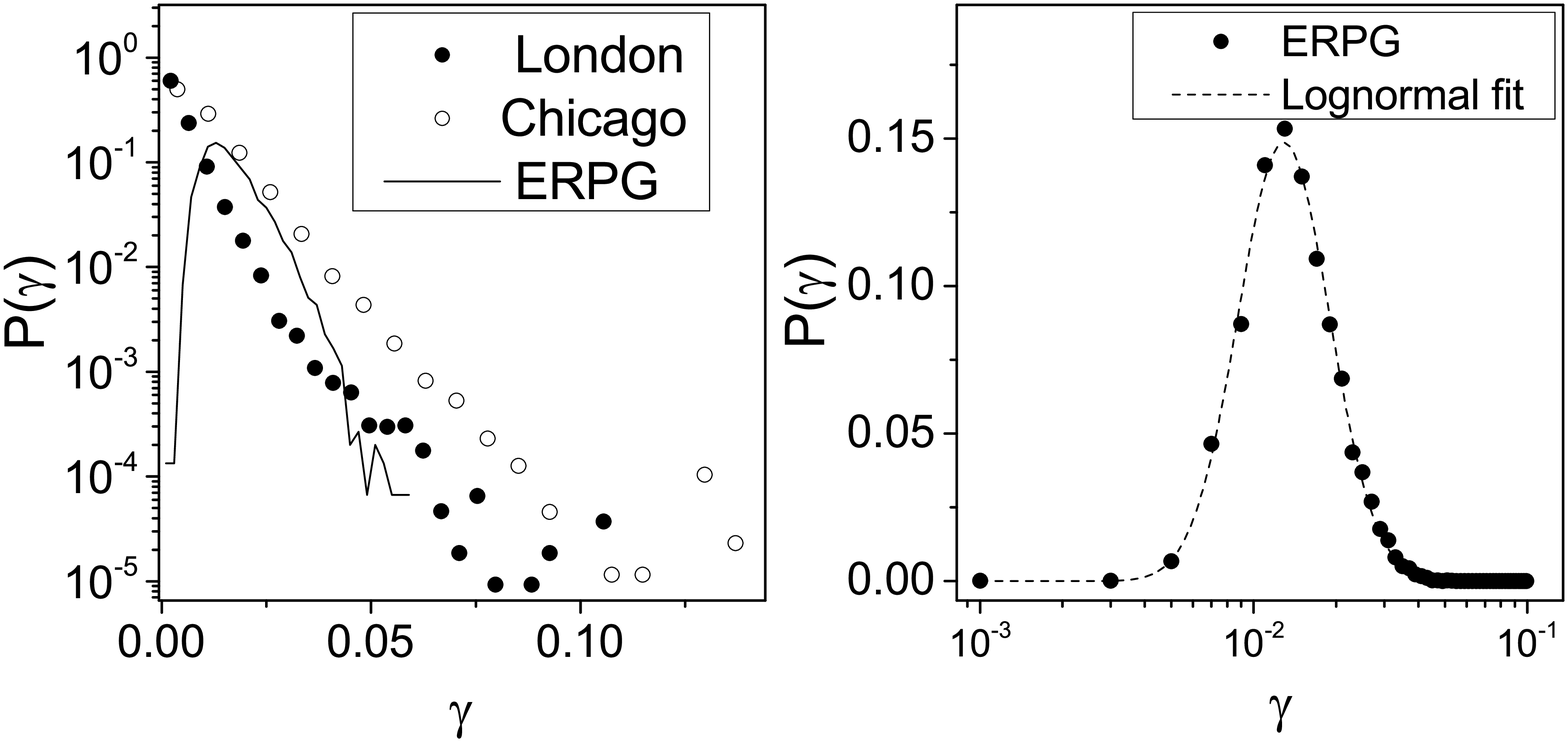}
	\caption{Top-left panel: Information content distribution $P(IC)$ for London, Chicago and ERPG. The dashed line is the exponential fit for the ERPG distribution. Top-right panel: normalized Shannon entropy $S(IC)$ calculated for the information content distribution for London, Chicago, ERPG and GRID. Bottom panels: $\gamma$ probability distribution $P(\gamma)$ for London, Chicago and ERPG (left), and ERPG (right). }
	\label{fig5}
\end{center} 
\end{figure}

In order to better understand how the information space affects the navigation in the primal space, first we consider a simple measure for the information content $IC$ of the nodes in the dual space, where $IC$ is defined as the number of street segments belonging to the same node in the dual space.
We expect this measure to give us some information on the information organization in the related transport systems.
In the left panel of Fig.\ref{fig5}, we show the probability distribution $P(IC)$ for the information content for London, Chicago and ERPG. 
The  distribution for the GRID is not shown, since it is trivial, i.e., all the nodes in the dual space represent the same number of street segments in the primal space.
We can observe how the distributions corresponding to the real networks are skewed in a log-log plot, revealing a hierarchical organization and the presence of hubs (notice that since the degree distribution in the primal space is nearly Poissonian, $P(IC)$ is nearly equivalent to the degree distribution in the dual space, see Fig.\ref{fig2}). 
On the other hand, $P(IC)$ for the ERPG is well fitted by an exponential function, revealing the lack of organization.

In the top-right panel of Fig.\ref{fig5}, we calculate the normalized Shannon entropy $S(IC)$ for the above mentioned distributions, i.e., $S\equiv\frac{-\sum P(IC)\ln(P(IC))}{N\ln N}$, where $N$ is the number of vertices in the network. 
$S(IC)$ is a measure of the unevenness of the related probability distributions.
It is $0$ for the GRID, it is maximum for the ERPG, while we find the real networks with intermediate values.

Finally, we introduce the \textit{informational closeness centrality} $C_{CL}^{Inf}$, another measure lying in the interplay between primal and dual representation, defined as the closeness centrality calculated in the primal graph, where the paths between the nodes are calculated along the shortest paths in the dual graph, i.e., $C_{CL}^{Inf}\equiv (N-1)/\sum_{i\neq j} d_{ij}^{Inf}$,  $d_{ij}^{Inf}$ is the shortest  Euclidean distance that connects node $i$ and node $j$ in the primal graph, calculated along the shortest path connecting node $i$ and $j$ in the dual space.

The informational closeness centrality is again a geometrical measure for the efficiency of the primal graph (how in average a node is distant from all other nodes of the graph), but incorporating the information driven by the dual space and we have that for each node $C_{CL}^{Inf} \geq C_{CL}$.

Then, for each node $i$ of the network, the ratio 
\begin{equation}
\gamma_i\equiv \frac{\frac{N-1}{\sum_j d_{ij}}-\frac{N-1}{\sum_j d_{ij}^{Inf}}}{\frac{N-1}{\sum_j d_{ij}}}=1-\frac{\sum_j d_{ij}}{\sum_j d_{ij}^{Inf}},
\end{equation}
 is a measure of the influence of the dual space for the transportation properties for that node. 
We have that $0\leq\gamma< 1$, where $\gamma=0$ if the shortest paths in the primal network are equivalent to the ones in the dual network, which is the case of a grid where all the shortest paths from a point to another one have the same length as the path along the shortest path in the dual graph, or as for a minimum spanning tree, where there is only one shortest path between each pairs of nodes.
Small values for $gamma$ mean that shortest paths in the primal space are closely following shortest paths in the dual space, while larger values for $gamma$ represent the opposite behaviour, where the information space topological behaviour is more influential in the street network navigation process.

In the bottom panels of Fig.\ref{fig5}, we show $P(\gamma)$ for London, Chicago and ERPG. 
The distribution for the GRID is trivial, as $\gamma=0$ for each node.
In the right panel we can see that for the ERPG $P(\gamma)$ is well fitted by a lognormal distribution.
For the real street networks this is not the case, as we find skewed distributions in a semi-log plane. 
In particular we find that for Chicago $P(\gamma)$ is broader than it is for London. 
This means that in average, in terms of shortest paths connecting different points in the network, the information space for Chicago is consistently more influential than the London's one and thus confirms what we previously find for the informational robustness analysis.

\section{Conclusions}

In this paper we analysed two relevant properties of street networks, the robustness and the closeness centrality.
In particular, we took as a sample two very large metropolis, London and Chicago.
While London is of Roman origin and reflects a long line of urban evolution spanning over two millennia, Chicago is the result of an urban explosion during the latter half of the nineteenth and the twentieth centuries. 
In this sense, our analysis is important to understand the physical properties of two different urbanization processes, the so-called self-organized and the planned one.
Worthless to say, this is a case study and we would not expect that all planned and self-organised cities behave in the same way. 
Nevertheless important conclusions could be extrapolated.

Our informational robustness analysis is novel, as it relates to the interplay between the primal and dual space, understanding the dual space as the main topological space shaping the street network as a transportation system.
Its relationship with the classical robustness analysis allows us for a swift results interpretation.
Moreover, by introducing two null models, the Delaunay triangulation and the minimum spanning tree, we are able to frame and give a correct interpretation for the closeness centrality results.
Finally, we are able to interpret the robustness results by introducing the novel coefficient $\gamma$, as the ratio between the closeness centrality and the novel informational closeness centrality.

On a first instance, we find that in terms of classical robustness, London is a strikingly fragile systems, with respect to Chicago and to the ERPG and GRID models.
Nevertheless, when we calculate the informational robustness, we see that London and Chicago display a very close fracture point.
This result lets us wonder on the fact that even if in the primal space the networks are very distinct in terms of morphology and topological properties, the interplay with the dual space creates two systems which are surprisingly similar in terms of transportation properties (see also the degree distributions in Fig.\ref{fig2}).

With the closeness distribution analysis, we then show that the high fragility of the real networks with respect to the models could be explained by the fact that such a fragility is required to ensure high performances of the transportation system. 
In this sense, we show that the real primal street networks in consideration perform nearly as well as their Delaunay triangulation, in terms of their closeness centrality, with just the $36\%-40\%$ of the total street length of the latter one. 

We finally notice, through the informational closeness analysis, that the dual representation for  Chicago's street network is more influential than that of London in terms of street network navigation.
 
We believe that the results shown in this paper could be relevant for a better understanding of urban systems.
The similarities we find between so different metropolis induce us to believe in the existence of common principles behind the organization of such complex systems.   
Such understanding could be helpful both in terms of the study of self-organizing systems and for urban policy making.

As we noticed in the introduction, street networks are not the only systems in nature whose properties could be better understood in the interplay between a physical and a virtual layer. 
This induces us to believe that models of content based networks, such as the one reported in \cite{mungan2005analytical}, or tensorial representations of multiplex networks \cite{de2013mathematical} would be of great benefit for a better understanding of such phenomena.

\begin{acknowledgments}
APM and CM were partially funded by the MECHANICITY Project (249393 ERC-2009-AdG). We would like to thank Dr. Elsa Arcaute for useful discussions and support.
\end{acknowledgments} 

\bibliography{lond}

\providecommand{\noopsort}[1]{}\providecommand{\singleletter}[1]{#1}%
\begin{thebibliography}{35}
\expandafter\ifx\csname natexlab\endcsname\relax\def\natexlab#1{#1}\fi
\expandafter\ifx\csname bibnamefont\endcsname\relax
  \def\bibnamefont#1{#1}\fi
\expandafter\ifx\csname bibfnamefont\endcsname\relax
  \def\bibfnamefont#1{#1}\fi
\expandafter\ifx\csname citenamefont\endcsname\relax
  \def\citenamefont#1{#1}\fi
\expandafter\ifx\csname url\endcsname\relax
  \def\url#1{\texttt{#1}}\fi
\expandafter\ifx\csname urlprefix\endcsname\relax\def\urlprefix{URL }\fi
\providecommand{\bibinfo}[2]{#2}
\providecommand{\eprint}[2][]{\url{#2}}

\bibitem[{\citenamefont{Blinder et~al.}(2010)\citenamefont{Blinder, Shih,
  Rafie, and Kleinfeld}}]{klein}
\bibinfo{author}{\bibfnamefont{P.}~\bibnamefont{Blinder}},
  \bibinfo{author}{\bibfnamefont{A.}~\bibnamefont{Shih}},
  \bibinfo{author}{\bibfnamefont{C.}~\bibnamefont{Rafie}}, \bibnamefont{and}
  \bibinfo{author}{\bibfnamefont{D.}~\bibnamefont{Kleinfeld}},
  \bibinfo{journal}{Proc. Natl. Acad. Sci. USA} \textbf{\bibinfo{volume}{107}},
  \bibinfo{pages}{12670} (\bibinfo{year}{2010}).

\bibitem[{\citenamefont{Gallos et~al.}(2012)\citenamefont{Gallos, Makse, and
  Sigman}}]{gallos2012small}
\bibinfo{author}{\bibfnamefont{L.~K.} \bibnamefont{Gallos}},
  \bibinfo{author}{\bibfnamefont{H.~A.} \bibnamefont{Makse}}, \bibnamefont{and}
  \bibinfo{author}{\bibfnamefont{M.}~\bibnamefont{Sigman}},
  \bibinfo{journal}{Proceedings of the National Academy of Sciences}
  \textbf{\bibinfo{volume}{109}}, \bibinfo{pages}{2825} (\bibinfo{year}{2012}).

\bibitem[{\citenamefont{Theraulaz et~al.}(2002)\citenamefont{Theraulaz,
  Bonabeau, Nicolis, Sol\'e, Fourcassie, Blanco, Fournier, Joly, Fern\'andez,
  Grimal et~al.}}]{ants}
\bibinfo{author}{\bibfnamefont{G.}~\bibnamefont{Theraulaz}},
  \bibinfo{author}{\bibfnamefont{B.}~\bibnamefont{Bonabeau}},
  \bibinfo{author}{\bibfnamefont{S.}~\bibnamefont{Nicolis}},
  \bibinfo{author}{\bibfnamefont{R.}~\bibnamefont{Sol\'e}},
  \bibinfo{author}{\bibfnamefont{V.}~\bibnamefont{Fourcassie}},
  \bibinfo{author}{\bibfnamefont{S.}~\bibnamefont{Blanco}},
  \bibinfo{author}{\bibfnamefont{R.}~\bibnamefont{Fournier}},
  \bibinfo{author}{\bibfnamefont{J.}~\bibnamefont{Joly}},
  \bibinfo{author}{\bibfnamefont{P.}~\bibnamefont{Fern\'andez}},
  \bibinfo{author}{\bibfnamefont{A.}~\bibnamefont{Grimal}},
  \bibnamefont{et~al.}, \bibinfo{journal}{Proc. Natl. Acad. Sci. USA}
  \textbf{\bibinfo{volume}{99}}, \bibinfo{pages}{9645} (\bibinfo{year}{2002}).

\bibitem[{\citenamefont{Cardillo et~al.}(2013)\citenamefont{Cardillo, Zanin,
  G{\'o}mez-Garde{\~n}es, Romance, del Amo, and
  Boccaletti}}]{cardillo2012modeling}
\bibinfo{author}{\bibfnamefont{A.}~\bibnamefont{Cardillo}},
  \bibinfo{author}{\bibfnamefont{M.}~\bibnamefont{Zanin}},
  \bibinfo{author}{\bibfnamefont{J.}~\bibnamefont{G{\'o}mez-Garde{\~n}es}},
  \bibinfo{author}{\bibfnamefont{M.}~\bibnamefont{Romance}},
  \bibinfo{author}{\bibfnamefont{A.~J.~G.} \bibnamefont{del Amo}},
  \bibnamefont{and}
  \bibinfo{author}{\bibfnamefont{S.}~\bibnamefont{Boccaletti}},
  \bibinfo{journal}{Eur. Phys. J. Special Topics}
  \textbf{\bibinfo{volume}{215}}, \bibinfo{pages}{23} (\bibinfo{year}{2013}).

\bibitem[{\citenamefont{Batty}(2005)}]{mikecc}
\bibinfo{author}{\bibfnamefont{M.}~\bibnamefont{Batty}},
  \emph{\bibinfo{title}{Cities and Complexity}}, The MIT Press, Cambridge, MA
  (\bibinfo{year}{2005}).

\bibitem[{\citenamefont{Zipf}(1949)}]{zipf}
\bibinfo{author}{\bibfnamefont{G.}~\bibnamefont{Zipf}},
  \emph{\bibinfo{title}{Human behavior and the principle of least effort}},
  Addison-Wesley Press, Cambridge, MA (\bibinfo{year}{1949}).

\bibitem[{\citenamefont{Gibrat}(1931)}]{gibrat}
\bibinfo{author}{\bibfnamefont{R.}~\bibnamefont{Gibrat}},
  \emph{\bibinfo{title}{Les Inegalities Economiques}}, Sirey, Paris
  (\bibinfo{year}{1931}).

\bibitem[{\citenamefont{Batty and Longley}(2004)}]{mikefract}
\bibinfo{author}{\bibfnamefont{M.}~\bibnamefont{Batty}} \bibnamefont{and}
  \bibinfo{author}{\bibfnamefont{P.}~\bibnamefont{Longley}},
  \emph{\bibinfo{title}{Fractal Cities: A Geometry of Form and Function}},
  Academic Press, San Diego, CA and London (available at www.fractalcities.org
  ) (\bibinfo{year}{2004}).

\bibitem[{\citenamefont{Murcio et~al.}(2015)\citenamefont{Murcio, Masucci,
  Arcaute, and Batty}}]{murcio2015multifractal}
\bibinfo{author}{\bibfnamefont{R.}~\bibnamefont{Murcio}},
  \bibinfo{author}{\bibfnamefont{A.~P.} \bibnamefont{Masucci}},
  \bibinfo{author}{\bibfnamefont{E.}~\bibnamefont{Arcaute}}, \bibnamefont{and}
  \bibinfo{author}{\bibfnamefont{M.}~\bibnamefont{Batty}},
  \bibinfo{journal}{arXiv preprint arXiv:1505.02760}  (\bibinfo{year}{2015}).

\bibitem[{\citenamefont{Rosvall et~al.}(2005)\citenamefont{Rosvall, Trusina,
  Minnhagen, and Sneppen}}]{rosvaldual}
\bibinfo{author}{\bibfnamefont{M.}~\bibnamefont{Rosvall}},
  \bibinfo{author}{\bibfnamefont{A.}~\bibnamefont{Trusina}},
  \bibinfo{author}{\bibfnamefont{P.}~\bibnamefont{Minnhagen}},
  \bibnamefont{and} \bibinfo{author}{\bibfnamefont{K.}~\bibnamefont{Sneppen}},
  \bibinfo{journal}{Phys Rev Lett} \textbf{\bibinfo{volume}{94}}
  (\bibinfo{year}{2005}), ISSN \bibinfo{issn}{0031-9007}.

\bibitem[{\citenamefont{Porta et~al.}(2006)\citenamefont{Porta, Crucitti, and
  Latora}}]{portadual}
\bibinfo{author}{\bibfnamefont{S.}~\bibnamefont{Porta}},
  \bibinfo{author}{\bibfnamefont{P.}~\bibnamefont{Crucitti}}, \bibnamefont{and}
  \bibinfo{author}{\bibfnamefont{V.}~\bibnamefont{Latora}},
  \bibinfo{journal}{Phys A} \textbf{\bibinfo{volume}{369}},
  \bibinfo{pages}{853} (\bibinfo{year}{2006}), ISSN \bibinfo{issn}{0378-4371}.

\bibitem[{\citenamefont{Jiang and Claramunt}(2004)}]{jiangdual}
\bibinfo{author}{\bibfnamefont{B.}~\bibnamefont{Jiang}} \bibnamefont{and}
  \bibinfo{author}{\bibfnamefont{C.}~\bibnamefont{Claramunt}},
  \bibinfo{journal}{Environ Plann B} \textbf{\bibinfo{volume}{31}},
  \bibinfo{pages}{151} (\bibinfo{year}{2004}), ISSN \bibinfo{issn}{0265-8135}.

\bibitem[{\citenamefont{Masucci et~al.}(2014)\citenamefont{Masucci, Stanilov,
  and Batty}}]{pdual}
\bibinfo{author}{\bibfnamefont{A.~P.} \bibnamefont{Masucci}},
  \bibinfo{author}{\bibfnamefont{K.}~\bibnamefont{Stanilov}}, \bibnamefont{and}
  \bibinfo{author}{\bibfnamefont{M.}~\bibnamefont{Batty}},
  \bibinfo{journal}{Phys. Rev. E} \textbf{\bibinfo{volume}{89}},
  \bibinfo{pages}{012805} (\bibinfo{year}{2014}).

\bibitem[{\citenamefont{Masucci et~al.}(2009)\citenamefont{Masucci, Smith,
  Crooks, and Batty}}]{masuccilond}
\bibinfo{author}{\bibfnamefont{A.~P.} \bibnamefont{Masucci}},
  \bibinfo{author}{\bibfnamefont{D.}~\bibnamefont{Smith}},
  \bibinfo{author}{\bibfnamefont{A.}~\bibnamefont{Crooks}}, \bibnamefont{and}
  \bibinfo{author}{\bibfnamefont{M.}~\bibnamefont{Batty}},
  \bibinfo{journal}{Eur Phys J B} \textbf{\bibinfo{volume}{71}},
  \bibinfo{pages}{259} (\bibinfo{year}{2009}), ISSN \bibinfo{issn}{1434-6028}.

\bibitem[{\citenamefont{Batty}(2004)}]{batty2004new}
\bibinfo{author}{\bibfnamefont{M.}~\bibnamefont{Batty}}, \bibinfo{journal}{CASA
  Working Paper 75}  (\bibinfo{year}{2004}).

\bibitem[{\citenamefont{Hillier et~al.}(1976)\citenamefont{Hillier, Leaman,
  Stansall, and Bedford}}]{hillier1976space}
\bibinfo{author}{\bibfnamefont{B.}~\bibnamefont{Hillier}},
  \bibinfo{author}{\bibfnamefont{A.}~\bibnamefont{Leaman}},
  \bibinfo{author}{\bibfnamefont{P.}~\bibnamefont{Stansall}}, \bibnamefont{and}
  \bibinfo{author}{\bibfnamefont{M.}~\bibnamefont{Bedford}},
  \bibinfo{journal}{Environment and Planning B: Planning and Design}
  \textbf{\bibinfo{volume}{3}}, \bibinfo{pages}{147} (\bibinfo{year}{1976}).

\bibitem[{\citenamefont{Hillier and Vaughan}(2007)}]{hillier2007city}
\bibinfo{author}{\bibfnamefont{B.}~\bibnamefont{Hillier}} \bibnamefont{and}
  \bibinfo{author}{\bibfnamefont{L.}~\bibnamefont{Vaughan}},
  \bibinfo{journal}{Progress in Planning} \textbf{\bibinfo{volume}{67}},
  \bibinfo{pages}{205} (\bibinfo{year}{2007}).

\bibitem[{\citenamefont{Xie and Levinson}(2007)}]{xie2007measuring}
\bibinfo{author}{\bibfnamefont{F.}~\bibnamefont{Xie}} \bibnamefont{and}
  \bibinfo{author}{\bibfnamefont{D.}~\bibnamefont{Levinson}},
  \bibinfo{journal}{Geographical analysis} \textbf{\bibinfo{volume}{39}},
  \bibinfo{pages}{336} (\bibinfo{year}{2007}).

\bibitem[{\citenamefont{Dorogovtsev and
  Mendes}(2001)}]{dorogovtsev2001language}
\bibinfo{author}{\bibfnamefont{S.~N.} \bibnamefont{Dorogovtsev}}
  \bibnamefont{and} \bibinfo{author}{\bibfnamefont{J.~F.~F.}
  \bibnamefont{Mendes}}, \bibinfo{journal}{Proceedings of the Royal Society of
  London. Series B: Biological Sciences} \textbf{\bibinfo{volume}{268}},
  \bibinfo{pages}{2603} (\bibinfo{year}{2001}).

\bibitem[{\citenamefont{Masucci et~al.}(2011)\citenamefont{Masucci, Kalampokis,
  Egu{\'\i}luz, and Hern{\'a}ndez-Garc{\'\i}a}}]{masucci2011wikipedia}
\bibinfo{author}{\bibfnamefont{A.~P.} \bibnamefont{Masucci}},
  \bibinfo{author}{\bibfnamefont{A.}~\bibnamefont{Kalampokis}},
  \bibinfo{author}{\bibfnamefont{V.~M.} \bibnamefont{Egu{\'\i}luz}},
  \bibnamefont{and}
  \bibinfo{author}{\bibfnamefont{E.}~\bibnamefont{Hern{\'a}ndez-Garc{\'\i}a}},
  \bibinfo{journal}{PloS one} \textbf{\bibinfo{volume}{6}},
  \bibinfo{pages}{e17333} (\bibinfo{year}{2011}).

\bibitem[{\citenamefont{Viana et~al.}(2013)\citenamefont{Viana, Strano, Bordin,
  and Barthelemy}}]{viana2013simplicity}
\bibinfo{author}{\bibfnamefont{M.~P.} \bibnamefont{Viana}},
  \bibinfo{author}{\bibfnamefont{E.}~\bibnamefont{Strano}},
  \bibinfo{author}{\bibfnamefont{P.}~\bibnamefont{Bordin}}, \bibnamefont{and}
  \bibinfo{author}{\bibfnamefont{M.}~\bibnamefont{Barthelemy}},
  \bibinfo{journal}{Scientific reports} \textbf{\bibinfo{volume}{3}}
  (\bibinfo{year}{2013}).

\bibitem[{\citenamefont{LAGESSE et~al.}(2015)\citenamefont{LAGESSE, BORDIN, and
  DOUADY}}]{lagesse2015spatial}
\bibinfo{author}{\bibfnamefont{C.}~\bibnamefont{LAGESSE}},
  \bibinfo{author}{\bibfnamefont{P.}~\bibnamefont{BORDIN}}, \bibnamefont{and}
  \bibinfo{author}{\bibfnamefont{S.}~\bibnamefont{DOUADY}},
  \bibinfo{journal}{Network Science} \textbf{\bibinfo{volume}{3}},
  \bibinfo{pages}{156} (\bibinfo{year}{2015}).

\bibitem[{mer()}]{meridian}
\bibinfo{howpublished}{\url{http://www.ordnancesurvey.co.uk/}},
  \bibinfo{note}{last visited 05/2015}.

\bibitem[{ill()}]{illinois}
\bibinfo{howpublished}{\url{http://gis.dot.illinois.gov/gist2/}},
  \bibinfo{note}{last visited 05/2015}.

\bibitem[{\citenamefont{Masucci et~al.}(2015)\citenamefont{Masucci, Arcaute,
  Hatna, Stanilov, and Batty}}]{masucci2015logistic}
\bibinfo{author}{\bibfnamefont{A.~P.} \bibnamefont{Masucci}},
  \bibinfo{author}{\bibfnamefont{E.}~\bibnamefont{Arcaute}},
  \bibinfo{author}{\bibfnamefont{E.}~\bibnamefont{Hatna}},
  \bibinfo{author}{\bibfnamefont{K.}~\bibnamefont{Stanilov}}, \bibnamefont{and}
  \bibinfo{author}{\bibfnamefont{M.}~\bibnamefont{Batty}},
  \bibinfo{journal}{Journal of The Royal Society Interface}
  \textbf{\bibinfo{volume}{12}}, \bibinfo{pages}{20150763}
  (\bibinfo{year}{2015}).

\bibitem[{\citenamefont{Kalapala et~al.}(2006)\citenamefont{Kalapala,
  Sanwalani, Clauset, and Moore}}]{kalapala2006scale}
\bibinfo{author}{\bibfnamefont{V.}~\bibnamefont{Kalapala}},
  \bibinfo{author}{\bibfnamefont{V.}~\bibnamefont{Sanwalani}},
  \bibinfo{author}{\bibfnamefont{A.}~\bibnamefont{Clauset}}, \bibnamefont{and}
  \bibinfo{author}{\bibfnamefont{C.}~\bibnamefont{Moore}},
  \bibinfo{journal}{Physical Review E} \textbf{\bibinfo{volume}{73}},
  \bibinfo{pages}{026130} (\bibinfo{year}{2006}).

\bibitem[{\citenamefont{Watts and Strogatz}(1998)}]{watts1998collective}
\bibinfo{author}{\bibfnamefont{D.~J.} \bibnamefont{Watts}} \bibnamefont{and}
  \bibinfo{author}{\bibfnamefont{S.~H.} \bibnamefont{Strogatz}},
  \bibinfo{journal}{nature} \textbf{\bibinfo{volume}{393}},
  \bibinfo{pages}{440} (\bibinfo{year}{1998}).

\bibitem[{\citenamefont{Dorogovtsev and
  Mendes}(2013)}]{dorogovtsev2013evolution}
\bibinfo{author}{\bibfnamefont{S.~N.} \bibnamefont{Dorogovtsev}}
  \bibnamefont{and} \bibinfo{author}{\bibfnamefont{J.~F.}
  \bibnamefont{Mendes}}, \emph{\bibinfo{title}{Evolution of networks: From
  biological nets to the Internet and WWW}} (\bibinfo{publisher}{Oxford
  University Press}, \bibinfo{year}{2013}).

\bibitem[{\citenamefont{Buhl et~al.}(2006)\citenamefont{Buhl, Gautrais, Reeves,
  Sol{\'e}, Valverde, Kuntz, and Theraulaz}}]{buhl2006topological}
\bibinfo{author}{\bibfnamefont{J.}~\bibnamefont{Buhl}},
  \bibinfo{author}{\bibfnamefont{J.}~\bibnamefont{Gautrais}},
  \bibinfo{author}{\bibfnamefont{N.}~\bibnamefont{Reeves}},
  \bibinfo{author}{\bibfnamefont{R.}~\bibnamefont{Sol{\'e}}},
  \bibinfo{author}{\bibfnamefont{S.}~\bibnamefont{Valverde}},
  \bibinfo{author}{\bibfnamefont{P.}~\bibnamefont{Kuntz}}, \bibnamefont{and}
  \bibinfo{author}{\bibfnamefont{G.}~\bibnamefont{Theraulaz}},
  \bibinfo{journal}{The European Physical Journal B-Condensed Matter and
  Complex Systems} \textbf{\bibinfo{volume}{49}}, \bibinfo{pages}{513}
  (\bibinfo{year}{2006}).

\bibitem[{\citenamefont{Stauffer and Aharony}(1985)}]{Stauffer}
\bibinfo{author}{\bibfnamefont{D.}~\bibnamefont{Stauffer}} \bibnamefont{and}
  \bibinfo{author}{\bibfnamefont{A.}~\bibnamefont{Aharony}},
  \emph{\bibinfo{title}{Introduction To Percolation Theory}}, Taylor and
  Francis Ltd., London (\bibinfo{year}{1985}).

\bibitem[{\citenamefont{Masucci et~al.}(2013)\citenamefont{Masucci, Stanilov,
  and Batty}}]{masuccipo}
\bibinfo{author}{\bibfnamefont{A.}~\bibnamefont{Masucci}},
  \bibinfo{author}{\bibfnamefont{K.}~\bibnamefont{Stanilov}}, \bibnamefont{and}
  \bibinfo{author}{\bibfnamefont{M.}~\bibnamefont{Batty}},
  \bibinfo{journal}{PLoS ONE} \textbf{\bibinfo{volume}{8}},
  \bibinfo{pages}{e69469} (\bibinfo{year}{2013}).

\bibitem[{\citenamefont{Mungan et~al.}(2005)\citenamefont{Mungan,
  Kabaklo{\u{g}}lu, Balcan, and Erzan}}]{mungan2005analytical}
\bibinfo{author}{\bibfnamefont{M.}~\bibnamefont{Mungan}},
  \bibinfo{author}{\bibfnamefont{A.}~\bibnamefont{Kabaklo{\u{g}}lu}},
  \bibinfo{author}{\bibfnamefont{D.}~\bibnamefont{Balcan}}, \bibnamefont{and}
  \bibinfo{author}{\bibfnamefont{A.}~\bibnamefont{Erzan}},
  \bibinfo{journal}{Journal of Physics A: Mathematical and General}
  \textbf{\bibinfo{volume}{38}}, \bibinfo{pages}{9599} (\bibinfo{year}{2005}).

\bibitem[{\citenamefont{De~Domenico et~al.}(2013)\citenamefont{De~Domenico,
  Sol{\'e}-Ribalta, Cozzo, Kivel{\"a}, Moreno, Porter, G{\'o}mez, and
  Arenas}}]{de2013mathematical}
\bibinfo{author}{\bibfnamefont{M.}~\bibnamefont{De~Domenico}},
  \bibinfo{author}{\bibfnamefont{A.}~\bibnamefont{Sol{\'e}-Ribalta}},
  \bibinfo{author}{\bibfnamefont{E.}~\bibnamefont{Cozzo}},
  \bibinfo{author}{\bibfnamefont{M.}~\bibnamefont{Kivel{\"a}}},
  \bibinfo{author}{\bibfnamefont{Y.}~\bibnamefont{Moreno}},
  \bibinfo{author}{\bibfnamefont{M.~A.} \bibnamefont{Porter}},
  \bibinfo{author}{\bibfnamefont{S.}~\bibnamefont{G{\'o}mez}},
  \bibnamefont{and} \bibinfo{author}{\bibfnamefont{A.}~\bibnamefont{Arenas}},
  \bibinfo{journal}{Physical Review X} \textbf{\bibinfo{volume}{3}},
  \bibinfo{pages}{041022} (\bibinfo{year}{2013}).

\bibitem[{\citenamefont{Arcaute et~al.}(2015)\citenamefont{Arcaute, Hatna,
  Ferguson, Youn, Johansson, and Batty}}]{arcaute2015constructing}
\bibinfo{author}{\bibfnamefont{E.}~\bibnamefont{Arcaute}},
  \bibinfo{author}{\bibfnamefont{E.}~\bibnamefont{Hatna}},
  \bibinfo{author}{\bibfnamefont{P.}~\bibnamefont{Ferguson}},
  \bibinfo{author}{\bibfnamefont{H.}~\bibnamefont{Youn}},
  \bibinfo{author}{\bibfnamefont{A.}~\bibnamefont{Johansson}},
  \bibnamefont{and} \bibinfo{author}{\bibfnamefont{M.}~\bibnamefont{Batty}},
  \bibinfo{journal}{Journal of The Royal Society Interface}
  \textbf{\bibinfo{volume}{12}}, \bibinfo{pages}{20140745}
  (\bibinfo{year}{2015}).

\bibitem[{\citenamefont{Jiang and Jia}(2011)}]{jiang2011zipf}
\bibinfo{author}{\bibfnamefont{B.}~\bibnamefont{Jiang}} \bibnamefont{and}
  \bibinfo{author}{\bibfnamefont{T.}~\bibnamefont{Jia}},
  \bibinfo{journal}{International Journal of Geographical Information Science}
  \textbf{\bibinfo{volume}{25}}, \bibinfo{pages}{1269} (\bibinfo{year}{2011}).

\end{thebibliography}

\end{document}